\RequirePackage{fix-cm}
\documentclass[]{svjour3}                     

\smartqed  

\usepackage{graphicx}
\usepackage{amsmath}
\usepackage{mathrsfs}
\usepackage{cite}
\usepackage{verbatim}
\usepackage{marvosym}
\usepackage{latexsym}
\usepackage{lineno,hyperref}
\usepackage{amssymb}
\usepackage[noend]{algpseudocode}
\usepackage{algorithmicx,algorithm}
\usepackage{subfigure}
\usepackage{graphicx}  
\usepackage{float}  
\usepackage{subfigure}  
\begin{document}

\title{Quantum adversarial metric learning model based on triplet loss function  
}
\subtitle{}

\titlerunning{Quantum adversarial metric learning model based on triplet loss function}        

\author{Yan-Yan Hou    $^{\textbf{1,2}}$    \and
        Jian Li        $^{\textbf{3}}$        \and
        Xiu-Bo Chen    $^{\textbf{4,5}}$      \and
        Chong-Qiang Ye  $^{\textbf{1}}$        \and
}

\institute{
\Letter Jian Li\\
Lijian@bupt.edu.cn\\
$^{\textbf{1}}$School of Artificial Intelligence, Beijing University of Posts and Telecommunications, Beijing 100876, China.\\
$^{\textbf{2}}$College of Information Science and Engineering, ZaoZhuang University, ZaoZhuang Shandong 277160, China.\\
$^{\textbf{3}}$School of Cyberspace Security Security, Beijing University of Posts Telecommunications, Beijing 100876, China.\\
$^{\textbf{4}}$Information Security Center, State Key Laboratory of Networking and Switching Technology, Beijing University of Post and Telecommunications, Beijing 100876, China.\\
$^{\textbf{5}}$GuiZhou University, Guizhou Provincial Key Laboratory of Public Big Data, Guizhou Guiyang, 550025, China.\\
}

\date{Received: date / Accepted: date}
\maketitle
\begin{abstract}

Metric learning plays an essential role in image analysis and classification, and it has attracted more and more attention. In this paper, we propose a quantum adversarial metric learning (QAML) model based on the triplet loss function, where samples are embedded into the high-dimensional Hilbert space and the optimal metric is obtained by minimizing the triplet loss function. The QAML model employs entanglement and interference to build superposition states for triplet samples so that only one parameterized quantum circuit is needed to calculate sample distances, which reduces the demand for quantum resources. Considering the QAML model is fragile to adversarial attacks, an adversarial sample generation strategy is designed based on the quantum gradient ascent method, effectively improving the robustness against the functional adversarial attack. Simulation results show that the QAML model can effectively distinguish samples of MNIST and Iris datasets and has higher $\epsilon$-robustness accuracy over the general quantum metric learning. The QAML model is a fundamental research problem of machine learning. As a subroutine of classification and clustering tasks, the QAML model opens an avenue for exploring quantum advantages in machine learning.


\end{abstract}

\section{Introduction}
\label{intro}
Machine learning has developed rapidly in recent years and is widely used in artificial intelligence and big data fields. Quantum computing can efficiently process data in exponentially sizeable Hilbert space and is expected to achieve dramatic speedups in solving some classical computational problems. Quantum machine learning, as the interplay between machine learning and quantum physics, brings unprecedented promise to both disciplines. On the one hand, machine learning methods have been extended to quantum world and applied to the data analysis in quantum physics\cite{cong2019quantum}. On the other hand, quantum machine learning exploits quantum properties, such as entanglement and superposition, to revolutionize classical machine learning algorithms and achieves computational advantages over classical algorithms\cite{benedetti2019parameterized}. Metric Learning is the core problem of some machine learning tasks\cite{chen2018adversarial}, such as $k$-nearest neighbor, support vector machines, radial basis function networks, and $k$-means clustering. Its core work is to construct an appropriate distance metric that maximizes the similarities of samples of the same class and minimizes the similarities of samples from different classes. Linear and nonlinear methods can be used to implement metric learning. In general, linear models have a limited number of parameters and are unsuitable for characterizing high-order features of samples. Recently, neural networks have been adopted to establish nonlinear metric learning models, and promising results have been achieved in face recognition and feature matching.

Classical metric learning models usually extract low-dimensional representations of samples, which will lose some details of samples. Quantum states are in high-dimensional Hilbert spaces, and their dimensions grow exponentially with the number of qubits. This quantum enables quantum models to learn high-dimensional representations of samples without explicitly invoking a kernel function. A parameterized quantum circuit is used to map samples in high-dimensional Hilbert space. The optimal metric model is obtained by optimizing the loss function based on Hilbert-Schmidt distances. With the increase of the the dimension, this speed-up advantage will become more and more pronounced, and it is expected to achieve exponential growth in computing speeds.
In recent years, researchers began to study how to adopt quantum methods to implement metric learning. Lloyd\cite{lloyd2020quantum} firstly proposed a quantum metric learning model based on hybrid quantum-classical algorithms. A parameterized quantum circuit is used to map samples in high-dimensional Hilbert space. The optimal metric model is obtained by optimizing the loss function based on Hilbert-Schmidt distances. This model achieves better effects in classification tasks. Nhat\cite{nghiem2020unified} introduced quantum explicit and implicit metric learning approaches from the perspective of whether the target space is known or not. The research establishes the relationship between quantum metric learning and other quantum supervised learning models. The above two algorithms mainly focus on classification tasks. Metric learning is a fundamental problem in machine learning, which can be applied not only to classification but also to clustering, face recognition, and other issues. In our research, we are devoted to constructing a quantum metric learning model that can serve various machine learning tasks.

Angular distance is a vital metric that quantifies the included angle between normalized samples\cite{mao2019metric}. Angular distance focuses on the difference in the direction of samples and is more robust to the variation of local feature\cite{wang2017deep},\cite{duan2018deep}. Considering the similarities between angular distances of classical data and inner products of quantum states, we design a quantum adversarial metric learning (QAML) model based on inner product distances, which is more suitable for image-related tasks. Unlike other quantum metric learning models, the QAML model maps samples from different classes into quantum superposition states and utilizes simple interface circuits to compute metric distances for multiple sample pairs in parallel. Furthermore, quantum systems in high-dimensional Hilbert space have counter-intuitive geometrical properties\cite{liu2020vulnerability}. The QAML model using only natural samples is vulnerable to adversarial attacks, under which some samples are closer to the false class, so the model is easy to make wrong judgements\cite{madry2017towards}. To solve this issue, we construct adversarial samples based on natural samples. The model's robustness is improved by the alternative train of natural and adversarial samples.
Our work has two main contributions:(i) We explore a quantum method to compute the triplet loss function, which utilizes quantum superposition states to calculate sample distances in parallel and reduce the demand for quantum resources. (ii) We design an adversarial samples generation strategy based on the quantum gradient ascent, and the robustness of the QAML model is significantly improved by alternatively training generated adversarial samples and natural samples. Simulation results show that the QAML model separates samples by a larger margin and has better robustness for functional adversarial attacks than general quantum metric learning models.

The paper is organized as follows. Section 2 gives the basic method of the QAML model. Section 3 verifies the performances of the QAML model. Finally, we get a conclusion and discuss the future research directions.

\section{Quantum adversarial metric learning}
\subsection{Preliminary theory}
Triplet loss function is a widely used strategy for metric learning\cite{salakhutdinov2007learning}, commonly used in image retrieval and face recognition. A triplet set ${(x_i^a, x^p_i, x^n_i)}$ consists of three samples from two classes, where anchor sample $x_i^a$ and positive sample $x^p_i$ belong to the same class, and negative sample $x^n_i$ comes from another class. The goal of metric learning based on triplet loss function is to find the optimal embedded representation space, in which positive sample pairs ${(x_i^a, x^p_i)}$ are pulled together and negative sample pairs ${(x_i^a, x^n_i)}$ are pushed away. Fig.1 shows sample space change in the metric learning process. As we can see, samples from different classes become linearly separable through metric learning. Fig.2 shows the schematic of the metric learning model based on triplet loss function. Firstly, the model prepares multiple triplet sets, and one triplet set ${(x_i^a, x^p_i, x^n_i)}$ is sent to convolutional neural networks (CNN), where three CNN with the same structure and parameters are needed. Each CNN acts on one sample of the triplet set to extract its features. The triplet loss function is obtained by computing metric distances for multiple sample pairs of triplet sets. In the learning process, the optimal parameters of CNN are obtained by minimizing the triplet loss function.
\begin{figure*}[t]
\centering
\includegraphics[scale=0.6]{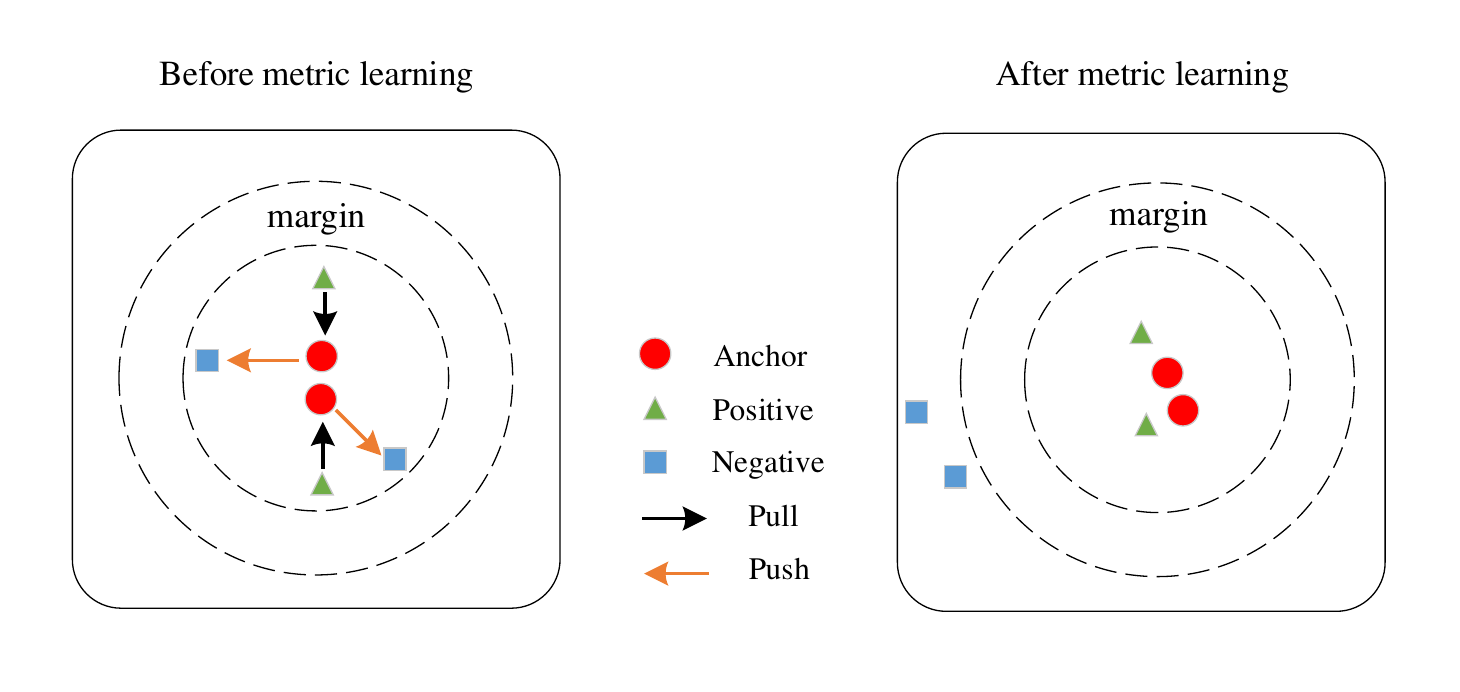}
\vspace{0cm}
\caption{Sample space change in metric learning process. Before metric learning, the distances between negative sample pairs are smaller, and samples from different classes are difficult to separate by linear functions. After metric learning, the distances between negative sample pairs become larger, and a large margin separates samples from different classes. Linear functions can easily separate positive and negative samples.}
\label{figure2}
\vspace{0cm}
\end{figure*}
\begin{figure*}[t]
\centering
\includegraphics[scale=0.6]{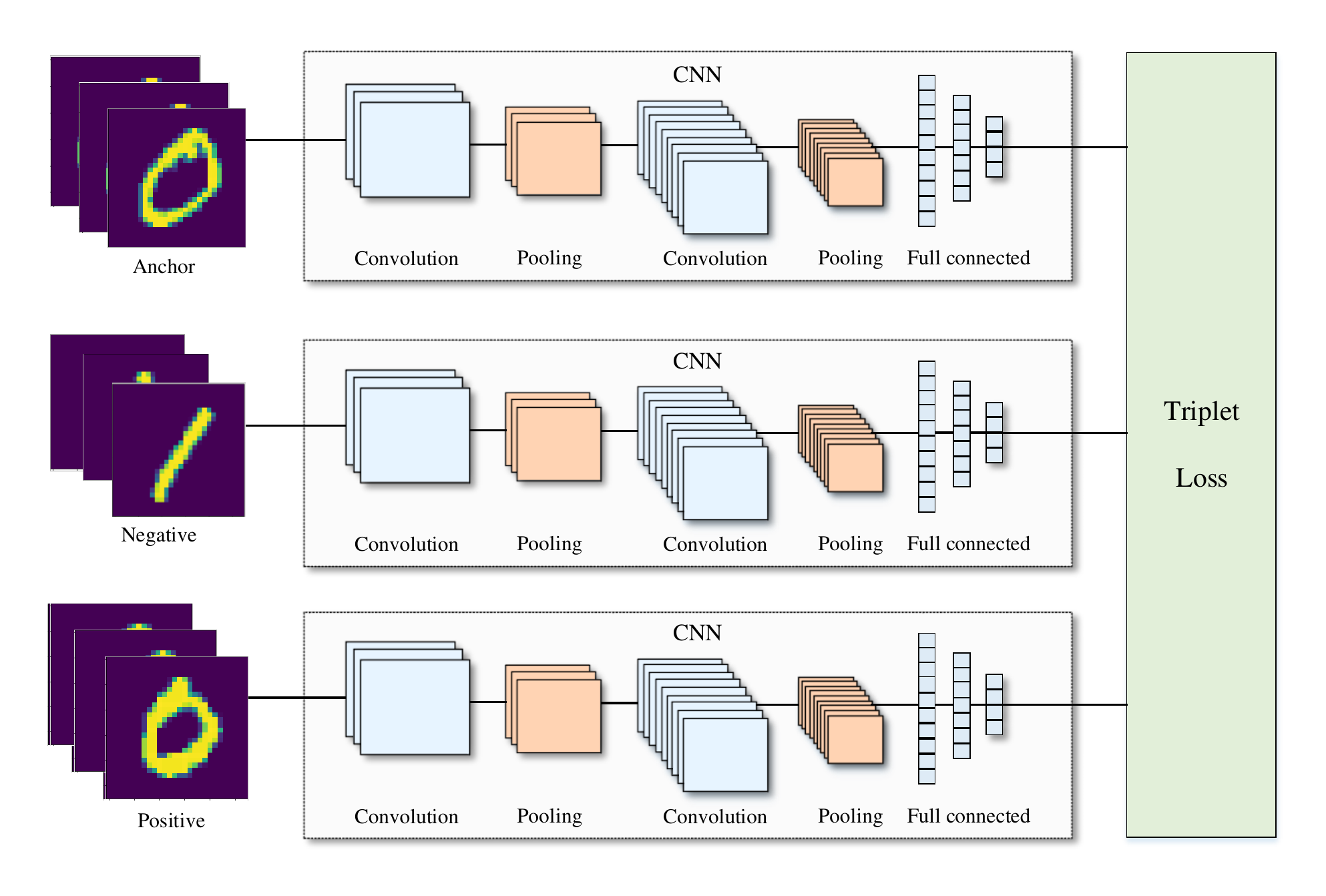}
\vspace{0cm}
\caption{The schematic of the metric learning model based on triplet loss function. A triplet set includes an anchor sample, a positive sample, and a negative sample. The input consists of a batch of triplet sets, and only one triplet set serves as input in each iteration. Three CNN with the same structure and parameters are used to map the triplet set into the embedded representation space. CNN, consisting of multiple convolutions, pooling, and fully connected layers, is responsible for extracting the features of samples. The triplet loss function is further constructed based on the extracted features.}
\label{figure2}
\vspace{0cm}
\end{figure*}
Let one batch samples include $N_1$ triplet sets. The triplet loss function is
\begin{equation}
\label{E41}
\begin{aligned}
\begin{array}{ccc}
L= \frac{1}{N_1}\sum_{i=1}^{N_1}[D(g(x^a_i),g(x^p_i))-D(g(x^a_i),g(x^n_i))+\mu]_{+},
\end{array}
\end{aligned}
\end{equation}
where $g(\cdot)$ represents the function mapping input samples to the embedded representation space, $D(\cdot,\cdot)$ denotes the distance between a sample pair in the embedded representation space, and $[\ \cdot\ ,\ \cdot\ ]_{+}=max(0,\ \cdot\ )$ represents the hinge loss function. The goal of metric learning is to learn a metric that makes the distances between negative sample pairs greater than the distance between the corresponding positive sample pairs and satisfies the specified margin $\mu\in \mathbb{R}^+$\cite{mao2019metric}. In the triplet loss function, $D(g(x^a_i), g(x^p_i))$ penalizes the positive sample pair $(x^a_i, x^p_i)$ that is too far apart, and $D(g(x^a_i), g(x^n_i))$ penalizes the negative sample pair $(x^a_i,x^p_i)$ whose distance is less than the margin $\mu$.

Metric learning can adopt various distance metric methods. Angular distance metric is robust to image illumination and contrast variation \cite{wang2017deep}, which is an efficient way for metric learning tasks. In this method, samples need to be normalized to unit vectors in advance. The distance between a positive sample pair is
\begin{equation}
\label{E21}
\begin{aligned}
\begin{array}{ccc}
D(g(x^a_i), g(x^p_i))=1-\frac{|g(x^a_i)\cdot g(x^p_i)|}{||g(x^a_i)||_2||g(x^p_i)||_2},
\end{array}
\end{aligned}
\end{equation}
where $|\ |$ and $||\ ||_2$ represent $l_1$-norm and $l_2$-norm, respectively, and $\cdot$ denotes the inner product operation for two vectors. The distance between negative sample pairs can be calculated in the same way.

\subsection{Framework of quantum metric learning model}

For most machine learning tasks, it is often challenging to adopt simple linear functions to distinguish samples of different classes. According to kernel theory\cite{blank2020quantum}, samples in high-dimensional feature space have better distinguishability. Classical machine learning algorithms usually adopt kernel methods to map samples to high-dimensional feature space, where the mapped samples can be separated by simple linear functions. Quantum states with $n$-qubits are in $2^n$-dimensional Hilbert space, where quantum systems characterize the nonlinear features of data and efficiently process data through a series of linear unitary operations.

In the QAML model, samples should be firstly mapped into quantum systems by qubit encoding. The Hilbert space after encoding usually does not correspond to the optimal space for separating samples of different classes. To search for the optimal Hilbert space, the QAML model performs parameterized quantum circuits $W(\theta)$ on the encoded states\cite{grant2018hierarchical}. As different variable parameters $\theta$ correspond to different mapping spaces, we can search the optimal space by modifying parameters $\theta=(\theta_1^1,...,\theta_i^j)$. As long as $W(\theta)$ has strong expressivity, we can find the optimal Hilbert space by optimizing the loss function of metric learning\cite{perez2020data,schuld2021effect}. $W(\theta)$ with different structures and layers have different expressivity. The more layers $W(\theta)$ has, the stronger the expressivity, and the easier it is to find the optimal metric space.

The classical metric learning model based on triplet loss function requires three identical CNN to map triplet sets $(x_i^a, x_i^p, x_i^n)$ into the novel Hilbert space. To reduce the demand for quantum resources, we construct a quantum superposition state to represent one triplet set so that a triplet set only needs one $W(\theta)$ to transform it into Hilbert space. The core work of the building loss function is to compute inner products between sample pairs, but $W(\theta)$ and subsequent conjugate operation $W^{\dagger}(\theta)$ counteract each other's effects. To solve this issue, we add a repeated encoding operation after $W(\theta)$. It is worth mentioning that the repeated encoding operation is also conducive to the construction of high-dimensional features of samples.

The QAML model is mathematically represented as the minimization of the loss function with respect to the parameters $\theta$. The triplet loss function consists of metric distances for positive and negative sample pairs, so the kernel work of the QAML model is constructing the metric distances for sample pairs in the transformed Hilbert space. The mapping samples $h(x_i^a)/||h(x_i^a)||_2$ and $h(x_i^p)/||h(x_i^p)||_2$ of Equ.\ref{E21} are replaced by the quantum states of $x_i^a$ and $x_i^p$, then the second term of Equ.\ref{E21} is converted to the inner product between quantum states of the positive sample pair $(x_i^a, x_i^p)$, which can be got by the method of the Hadamard classifier\cite{blank2020quantum}. The triplet loss function can be viewed as the weighted sum of the inner product of sample pairs $(x_i^a, x_i^p)$ and the inner product of sample pairs $(x_i^a,x_i^n)$. With the help of ancilla registers, the triplet set can be prepared in superposition states form. According to the entanglement property of superposition states, the triplet loss function can be implemented with one parameterized quantum circuit. Then, the triplet loss function value is transmitted to a classical optimizer, and parameters are optimized until the optimal metric is obtained. The QAML model constructs adversarial samples according to the gradient of natural samples and trains alternatively natural and adversarial samples to improve the model's robustness against adversarial attacks. The schematic of the QAML model is shown in Fig.3.

\begin{figure*}[t]
\centering
\includegraphics[scale=0.5]{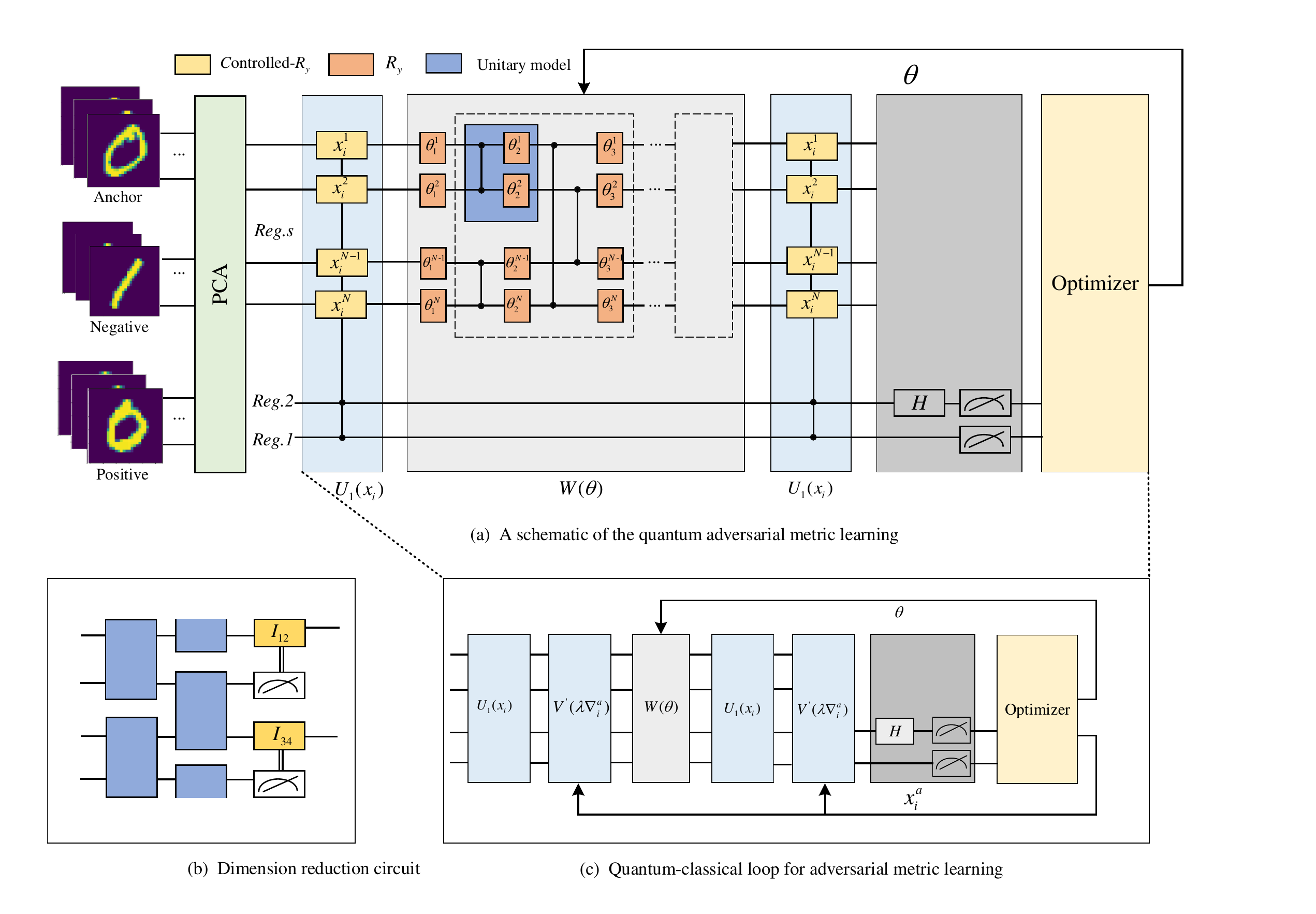}
\vspace{0cm}
\caption{Overview of quantum adversarial metric learning (QAML) model. Panel (a) shows the framework of quantum adversarial metric learning. $Reg.s$ is the sample register that stores triplet sets, and $Reg.1$ and $Reg.2$ are ancilla registers used distinguishing different samples. The model firstly adopts principal component analysis (PCA) to reduce the input dimension. Subsequently, anchor, negative and positive samples are encoded into a quantum superposition state by controlled qubit encoding. The transformation of Hilbert space is implemented by parameterized quantum circuit $W(\theta)$ and the subsequent qubit encoding $U_1(x_i)$. Finally, Hadamard and measurement operations act on ancilla registers to simultaneously compute the inner products for the positive and negative sample pairs, and the triplet loss function is further obtained. In each iteration, the parameters $\theta$ are updated by optimizing the triplet loss function with a classical optimizer. Panel (b) shows the quantum dimension reduction circuit to reduce the number of output qubits. In each module, only one qubit is measured, and the controlled unitary based on its measurement result acts on another qubit. Panel (c) shows another case of the QAML model, where adversarial samples are built and added to the training process. $V^{'}(\lambda\nabla_i^a)$ is the unitary operation based on the gradient of anchor sample $x_i^a$ and acts on the encoded quantum states to produce its adversarial sample. In the QAML model training process, natural and adversarial samples alternatively serve as input.}
\label{figure2}
\vspace{0cm}
\end{figure*}

\bibliographystyle{spphys}       
\bibliography{manuscript}   
\end{document}